\documentclass[english,twocolumn,showpacs,preprintnumbers,amsmath,amssymb,
showkeys]{revtex4}
\usepackage{graphicx}
\usepackage{dcolumn}
\usepackage{bm}
\usepackage{bigints}
\usepackage{babel,amssymb,amsmath,graphicx,hyperref}

  \DeclareMathOperator{\arccosh}{arccosh}

\begin{document}

\title{Disintegration and expansion of wormholes}

\author{I.D. Novikov}
\affiliation{Astro-Space Center of P.N. Lebedev Physical Institute, Profsoyusnaya 84/32, Moscow, Russia 117997,}
\affiliation{The Niels Bohr International Academy, The Niels Bohr Institute, Blegdamsvej 17, DK-2100, Copenhagen, Denmark}
\affiliation{National Research Center Kurchatov Institute, 1, Akademika Kurchatova pl., Moscow,  Russia 123182}
\author{D.I. Novikov}
\affiliation{Astro-Space Center of P.N. Lebedev Physical Institute, Profsoyusnaya 84/32, Moscow, Russia 117997}
\author{S.V. Repin}
\affiliation{Astro-Space Center of P.N. Lebedev Physical Institute, Profsoyusnaya 84/32, Moscow, Russia 117997}

\begin{abstract}
We consider the process of catastrophic expansion of a spacelike wormhole after a violation of its equilibrium state. 
The dynamics of deformation of the comoving reference frame is investigated. We show that the deformation has 
a very specific anisotropic feature. The statement made earlier by other authors, that in the process of expanding 
the wormhole connecting two universes these universes ultimately unite into one universe, is not correct. We show 
that the transverse size of the wormhole (its throat) increases and the length of the corridor decreases which 
does not correspond to the de Sitter model.
\end{abstract}

\maketitle

\section{Introduction}

      At present the very fact of the existence of wormholes (WHs) is
regarded as a purely theoretical hypothesis. However, there are very
characteristic features that make it possible to distinguish wormholes
from black holes.
In particular the special shape of the magnetic field in the
vicinity of the wormhole \cite{Novikov_2019a}, the presence of the blueshift due to matter
outflow  and visible structures within the shadow of the wormhole are
distinctive criteria of WHs. It should be noted that planned high-resolution
and high-sensitivity VLBI experiments such as the Origins Space Telescope
\cite{2018NatAs...2..596B}, Millimetron \cite{2014PhyU...57.1199K} and
the James Webb Space Telescope \cite{2018ConPh..59..251K}
can provide us (at least in principle) with the possibility of observing 
the features noted above which would indicate the existence of WHs. 
Therefore one can expect that in the foreseeable future the physics of wormholes 
from a purely theoretical science can become a part of observational astrophysics.

      One of the most important issues in the physics of wormholes 
is the problem of their evolution under the influence of certain circumstances
and especially the problem of their final state. WHs are the topological tunnels connecting
the distant regions of our Universe, or even different universes, with each other.
These objects are of different types \cite{Novikov_2018} and wormholes of
different types evolve in different ways. 
In Ref. \cite{Novikov_2019a} we considered the catastrophic collapse of
various types of WHs under the influence of a scalar field irradiation
with radiation density $\varepsilon > 0$. In Ref. \cite{Novikov_2019a}
we also considered the various final states the collapse of different types
of WHs can lead to.

      The evolution in the opposite direction is also of great interest, i.e. the expansion and 
breakup of a WH due to instability or under the action of a small perturbation by exotic matter 
with negative energy density~$\varepsilon <0$, \cite{Shinkai_2002,Doroshkevich_2009}. 
First of all, the classical spacelike WHs are of interest. In this article we study the process 
of wormhole expansion from the beginning to the final stage. The outline of our paper is 
the following: In Sec.~II we consider the initial stage of WH evolution after the loss
of equilibrium; in Sec.~III we demonstrate the results of
numerical analysis of the wormhole disintegration. In Sec.~IV (Discussion)
we discuss our results and compare them with some statements made earlier by
other authors. Finally, in Sec.~V we make our brief conclusions.
In the Appendix we separately discuss the homogeneous approximation.

\section{The initial stage of expansion}
\label{Initial_stage}

     Let us consider the initial stages of the collapse of the Ellis-Bronnikov-Morris-Thorne WH 
(Ref. \cite{Doroshkevich_2009}) after the loss of the equilibrium.
The spherical metric can be written as follows:
\begin{equation}
  \begin{array}{l}
    \vspace{0.5cm}  
  ds^2 = -d\tau^2 +e^\lambda dR^2 + r^2 d\Omega^2,
  r^2 = e^\eta(R^2 + Q^2),\\ 
  d\Omega^2 = d\vartheta^2 + \sin^2\vartheta\, d\varphi^2,
  \end{array}
\end{equation}
where $R$ is the radial coordinate and $Q$ is the throat size. Hereinafter
we use $c = 1$, $G = 1$.

    The equations for the scalar field $\Psi$ are
\begin{equation}
 \left.
  \begin{split}
     \left[\Psi,_{\tau}\hspace{0.1cm}e^{\eta + \lambda/2}\right]_{\text{\normalsize{,}}\tau}& =
     \frac{1}{Y^2}\left[Y^2\Psi,_{{}_{R}}\hspace{0.1cm}
       e^{\eta - \lambda/2}\right]_{\text{\normalsize{,}}R}\hspace{0.2cm}, \\[2mm]
       &Y^2 = R^2 + Q^2\,.
  \end{split}
 \right\}
\end{equation}
For a static field:
\begin{equation}
  \begin{array}{l}
    \vspace{0.5cm}
    \lambda = 0, \qquad \eta = 0, \qquad \Psi,_R = \pm\,\cfrac{Q}{Y^2}\,\,,\\
    \qquad
     T_\tau^\tau = -T_R^R = T_\vartheta^\vartheta = \cfrac{Q^2}{8\pi Y^4}\,\,.
  \end{array}
\end{equation}

     We will consider the evolution of the WH at the stage when $\lambda$ and $\eta$ are small in absolute value.
The perturbation equations for the linearized Einstein equations give for one of the solutions \cite{Doroshkevich_2009}:
\begin{eqnarray}
      \lambda & = & -2\eta\,, 
           \label{lambda_min_2_eta}   \\
      \eta_{,tt} & = & \eta_{,xx} + 
                                                                  \cfrac{6x}{1 + x^2}\,\,\eta_{,x} +
                                                                  \cfrac{6\eta}{1 + x^2}\,\,,
\end{eqnarray}
where $x = R/Q$ and  $t = \tau/Q$.

For $\eta \sim \exp(i\omega t)$ we have
\begin{equation}
  \begin{array}{l}
    \vspace{0.5cm}
    \eta = -\cfrac{\lambda}{2} = \cfrac{f}{\left(1+x^2\right)^{3/2}}
    \hspace{0.2cm},\\
  f,_{xx} + \left[\omega^2 + \cfrac{3}{\left(1 + x^2\right)^2} \right]f = 0.
  \end{array}
\end{equation}

      The analysis in Ref. \cite{Doroshkevich_2009} shows that for $|x| \ll 1$ the function $f$ grows and the central region 
of the wormhole around the throat $|R| < |xQ|$ is unstable: 
\begin{equation} 
  \begin{array}{l}
    \vspace{0.5cm}
       f \sim e^{\alpha t} \sin\left(x\sqrt{3 - \alpha^2}\right) + \Phi,\\ 
       \omega^2=-\alpha^2, \quad \alpha = \text{const}, \quad \Phi = \text{const}.
  \end{array}     
\end{equation}
For $|R| \ge |xQ|$ the function $f$ is exponentially small.

    We emphasize that Eq. (\ref{lambda_min_2_eta}) shows that if the expansion occurs in the direction
orthogonal  to the radial one, then the contraction occurs in the radial direction. We did not find another 
possible beginning of expansion with a qualitatively different deformation of the reference frame like in 
Eq.~(\ref{lambda_min_2_eta}).

\section{Numerical analysis of the disintegration of a wormhole}

      \label{Numeric_analysis}

      Let us consider numerically the disintegration process of a spherically symmetric WH without any additional 
restrictions. 

\begin{figure}[tbh] 
  \includegraphics[width=0.9\columnwidth]{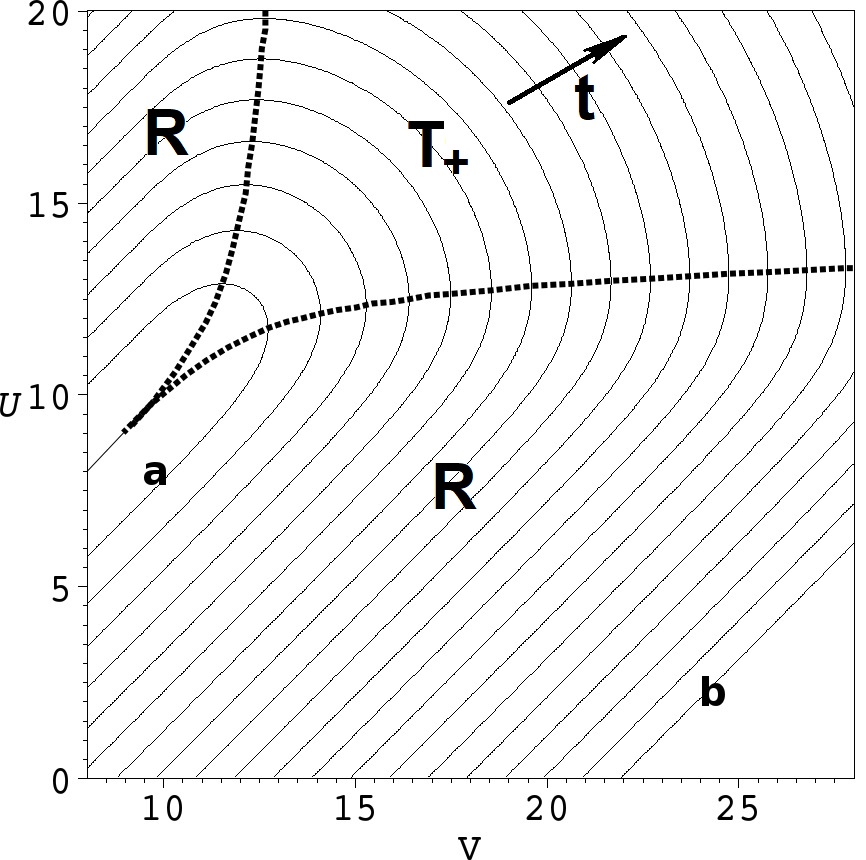}
  \includegraphics[width=0.9\columnwidth]{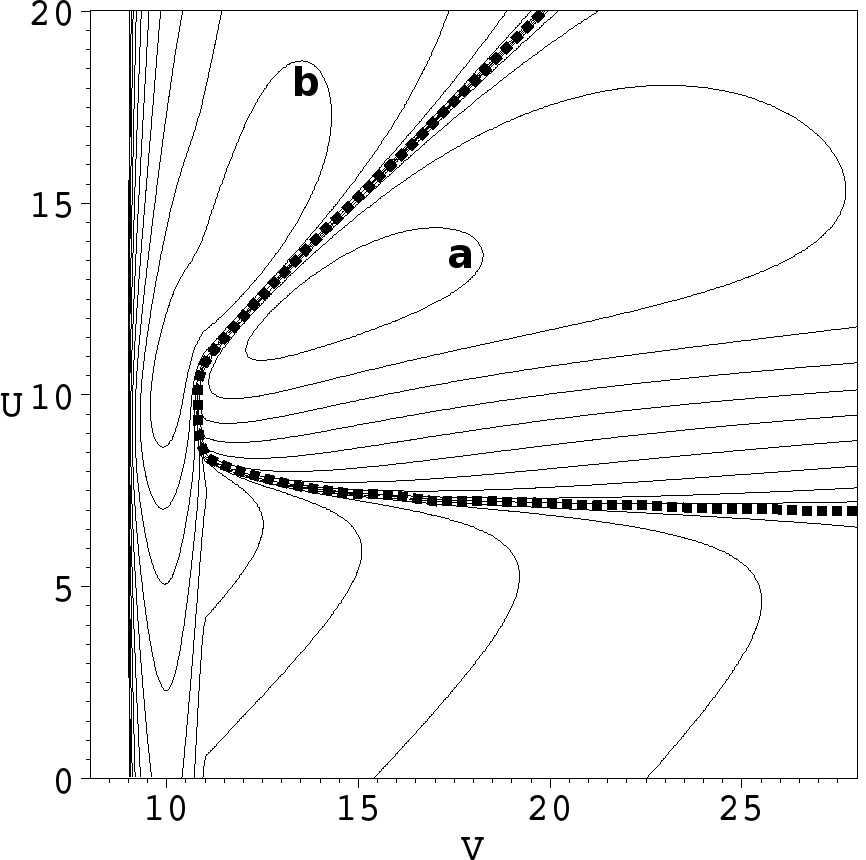} 
  \caption{Catastrophic expansion of WH$_{\text{MT}}$ after imbalance due to its irradiation from the side of 
                the $v$-coordinate by a short spherical pulse of a scalar field with $\varepsilon < 0$. See the text for 
                designations. Results of a simulation with $v_1 = 9$, $v_2 = 11$ and $A^\Psi = -0.01$. (a) Lines of 
                constant~$\tilde r$, from  $\tilde r = 1$ to $\tilde r = 5$ with~$\Delta \tilde r = 0.5$ between lines 
                (the line $\mathbf{a}$ marks $\tilde r = 1.5$ and the line $\mathbf{b}$ marks $\tilde r = 11$. $T_+$ 
                is an expanding $T$-region, $R$~is the $R$-region. Dotted lines
                mark apparent horizons. (b) Lines of constant $T_{vv}^\Psi - T_{uu}^\Psi$. 
                The region $\mathbf{a}$ is the region with $T_{vv}^\Psi - T_{uu}^\Psi < 0$, $\mathbf{b}$~marks 
                the region with $T_{vv}^\Psi - T_{uu}^\Psi > 0$ and the thick dotted line 
                marks $T_{vv}^\Psi - T_{uu}^\Psi = 0$.}
   \label{Figure_1}
\end{figure}

       Apparently, the first numerical simulation of nonlinear processes of the wormhole expansion was performed 
in Ref.~\cite{Shinkai_2002}. As we will show below, the authors of this paper do not agree with the conclusions 
of Ref.~\cite{Shinkai_2002} on the issue under consideration.

We will consider an expansion of a WH with a scalar exotic field $\Psi$ \cite{Doroshkevich_2009}.
A spherically symmetric metric can be written in double zero coordinates as
\begin{equation}
  \begin{array}{l}
    \vspace{0.5cm}
       ds^2 = -2e^{2b(u,v)}du\,dv + \tilde r^2(u,v)\, d\Omega^2\,\,,\\
                    d\Omega^2 = d\vartheta^2 + \sin^2\vartheta\,d\varphi^2\,.
                    \label{spher_sym_metric}
 \end{array}                   
\end{equation}
The energy-momentum tensor for the field $\Psi$ is written as
\begin{equation}
  \begin{array}{c}
    \vspace{0.5cm}
    T_{\mu\nu}^\Psi =-\cfrac{1}{4\pi}\times\\
    \times\begin{pmatrix}
       \Psi_{\text{\normalsize{,}}u}^2 & 0 &                                                  0 & 0 \\
       0 &                                                  \Psi_{\text{\normalsize{,}}v}^2 & 0 & 0 \\
       0 & 0 & \tilde r^2e^{-2\sigma}\Psi_{\text{\normalsize{,}}u} \Psi_{\text{\normalsize{,}}v} & 0 \\ 
       0 & 0 & 0 & \tilde r^2\sin^2\vartheta e^{-2\sigma}\Psi_{\text{\normalsize{,}}u} \Psi_{\text{\normalsize{,}}v}
    \end{pmatrix}.
    \end{array}
\end{equation}

\vspace{0.5cm}

The field $\Psi$ satisfies the Klein-Gordon equation:
\begin{equation}
      \nabla^\mu \nabla_\mu \Psi = 0\,.
      \label{nabla_nabla_eqn}
\end{equation}
The Einstein equations corresponding to Eqs. (\ref{spher_sym_metric})-(\ref{nabla_nabla_eqn}) can be 
found in Ref.\cite{Doroshkevich_2009}. We have solved these equations numerically using specially constructed 
code \cite{Doroshkevich_2009}. Our goal is to study the expansion of the Ellis-Bronnikov-Morris-Thorne WH 
(hereinafter we denote it as the WH$_{\text{MT}}$) starting from the static state. This state is irradiated from one 
hole by a narrow spherical impulse of a scalar field~$\Psi$.

     The static WH$_{\text{MT}}$ in $(u, v)$ coordinates is written as
\begin{equation} 
      \tilde r(u,v) = \sqrt{Q^2 + \cfrac 14\, (v - u)^2}\,\,, \qquad  -g_{uv} = 2,
      \label{WH_MT_r}
\end{equation}
\begin{equation}
      \Psi = \arctan\left(\cfrac{v - u}{2Q}\right) .
       \label{WH_MT_Psi}
\end{equation}
We will study the nonlinear processes of expansion of the WH $_{\text{MT}}$ Eqs. (\ref{WH_MT_r}) and (\ref{WH_MT_Psi}), 
arising when the equilibrium of this model is disturbed by the irradiation from the $v$-coordinate side by a spherical 
impulse of the exotic scalar field $\Psi$ with negative energy density. The numerical value of $Q$ in 
Eqs. (\ref{WH_MT_r})-(\ref{WH_MT_Psi}) is set equal to 1, $Q = 1$. The area of numerical simulation is limited 
by the values of the coordinates: $v = [8,\,28]$, $u = [0,\,20]$. The unknown functions which are to be determined 
in the simulation are $\tilde r(u, v)$, $\sigma(u, v)$ and $\Psi(u, v)$. The initial values of the impulse are set along 
$u = u_0 = 0$.

      The full initial value $\Psi(u_0, v)$ is given by the formula
\begin{equation}
       \Psi(u_0,v) = \Psi_{\text{MT}}(u_0,v) + \tilde\Psi(u_0,v)\,,
\end{equation}
where $\Psi_{\text{MT}}(u_0,v)$ is determined by Eq. (\ref{WH_MT_Psi}), and the impulses $\tilde\Psi(u_0,v)$ 
are determined by the expression
\begin{equation}
       \tilde\Psi_{\text{\normalsize{,}}v}(u_0,v) = A^\Psi \sin^2\left(\pi\left(\cfrac{v - v_2}{v_2 - v_1}\right)\right)\,.
\end{equation}
$A^\Psi$ is the impulse amplitude, $A^\Psi > 0$, and $v_2 - v_1$ is its width. Outside the segment $[v_1, v_2]$
the value~$\tilde\Psi_{\text{\normalsize{,}}v}$ is equal to zero. In the computations, the shape and the magnitude 
of the impulse have been varied. 

        Figures \ref{Figure_1}(a) and \ref{Figure_1}(b) represent the results of numerical simulation. 
In Fig.~\ref{Figure_1}(a) one can see the emergence of apparent horizons in the direction of the $u$- and 
$v$-coordinates, and the appearance of 
the corresponding $T_+$-region, which expands. The definition and description of the $R$- and $T$-regions one
can find in Refs.~\cite{Novikov_1962, Zeldovich_1967, Frolov_1998}. 

        For numerical simulation $v_1 = 9$, $v_2 = 11$, $A^\Psi = -0.01$ were chosen. The impulse first 
reaches the throat $\tilde r = 1$ at the moment $u_0 = v_0 = 9$. Such an impulse can be considered as a small 
initial perturbation of the equilibrium of  WH$_{\text{MT}}$. After this imbalance, the process develops in 
a nonlinear mode. 

      Figure \ref{Figure_1}(a) shows also the world lines $\tilde r = \text{const}$ of those points of the reference frame,
which are located along the radial coordinate. In the region $R$ these lines are timelike. Their behavior over time 
characterizes the deformation of the reference frame along the radial coordinate. In Fig.~\ref{Figure_1}(a) this 
behavior corresponds to compression along this coordinate.

      The lines of constant value $T_{vv}^\Psi - T_{uu}^\Psi$ of the scalar field $\Psi$ are shown in 
Fig.~\ref{Figure_1}(b). This difference describes the flux of the field in and out of the WH. It is this difference which
is important for dynamics. Figure~\ref{Figure_1}(b) shows the scattering of the $\tilde \Psi$ flux on the space-time 
curvature and the $\Psi$ field flux from the throat to the larger values of $\tilde r$. With the development of the process, 
the $\Psi$ field takes up more and more volume and flows into empty space through both exits from the wormhole.
In Ref. \cite{Novikov_2019a}, we considered the collapse of a wormhole. In this model, the outflow of the $\Psi$ field leads 
to an increase in the forces of attraction acting on the wormhole. As a result of this process, two black holes were formed 
with masses that were absent before the start of this process.

\section{Discussion}

     Let us return to the question posed at the beginning of the article. How does the WH disintegration occur after 
the loss of its staticity? This issue has been covered in Ref.\cite{Shinkai_2002}. The authors carried out a numerical 
study of the formation of the apparent horizon and the evolution of the WH throat and came to the following 
conclusion:
\begin{quote}
        ``This exponential expansion, combined with the horizon structure, indicates that the wormhole has exploded
        to an inflationary universe.''
        
        ``The two universes connected by the wormhole have essentially been combined into one universe.''
\end{quote}

Similar conclusions were made in Ref. \cite{Gonzalez_2009}:
\begin{quote}
         ``We show that small initial perturbation do not decay to zero but grow in time and eventually kick 
         the wormhole throat away from its equilibrium configuration.''
\end{quote}

       We disagree with this. The dynamics of the apparent horizon and the throat are insufficient for such conclusions.
Our numerical simulations show that the throat swells but does not disappear. The ``two universes'' in the $u$- 
and $v$- directions, connected by a throat, are preserved. From the very beginning of the emergence 
of dynamics and in the course of the evolution, the deformation of the comoving space occurs in a cardinally 
anisotropic manner. The compression occurs along the radial direction, and expansion occurs in the directions which 
are perpendicular to the radial one. We emphasize this in Sec.~\ref{Initial_stage} and in the Appendix, 
where the consideration was carried out in various approximations, and also in Sec.~\ref{Numeric_analysis}.
There is no any approximation to de Sitter's solution cited by the authors of \cite{Shinkai_2002}. In inflationary cosmological 
models, the isotropic expansion occurs in all directions.

      Let us compare the two models: the expanding WH and the expanding de Sitter model. We will take the identical 
static states as the initial states and investigate how they will evolve in the future. The evolution of a WH is 
described in Sec.~\ref{Numeric_analysis}. The evolution of the de~Sitter model is well known. For the stationary 
reference frame at the initial moment $t_0 = 0$ and later on reaching the state of the inflationary universe, 
see Ref.~\cite{Berezin_2008}. Correcting typos in Ref.~\cite{Berezin_2008} and using the $-+++$ signature we have
\begin{equation}
  \begin{array}{l}
    \vspace{0.5cm}
      ds^2 = -dt^2 + a_0^2\cosh^2\cfrac{t}{a_0} \left(d\chi^2 + \sin^2\chi\,d\Omega^2 \right),\\ 
      \qquad 0 \le t < \infty.
\end{array}      
          \label{DSframe}
\end{equation}
In the de Sitter universe, any point can be chosen as the center of spherical symmetry. In the reference 
frame Eq. \ref{DSframe} the equations for the apparent horizons are written as
\begin{equation}
       t = a_0 \arccosh\left(\cfrac{1}{a_0\sin\chi} \right)\,\,.
       \label{t_coord}
\end{equation}
Let us compare the deformations of the space of the WH reference frame (see Fig. \ref{Figure_1}) and 
the reference frame in Eq.~(\ref{DSframe}). We draw the lines of the constant Lagrangian radial coordinate.
 Such a coordinate can be a line of constant size of the WH radius at a given Lagrangian point, i.e. 
 the line $g_{22} = \text{const}$ in Eq. \ref{DSframe}.

\begin{figure}[tbh]
  \includegraphics[width=0.95\columnwidth]{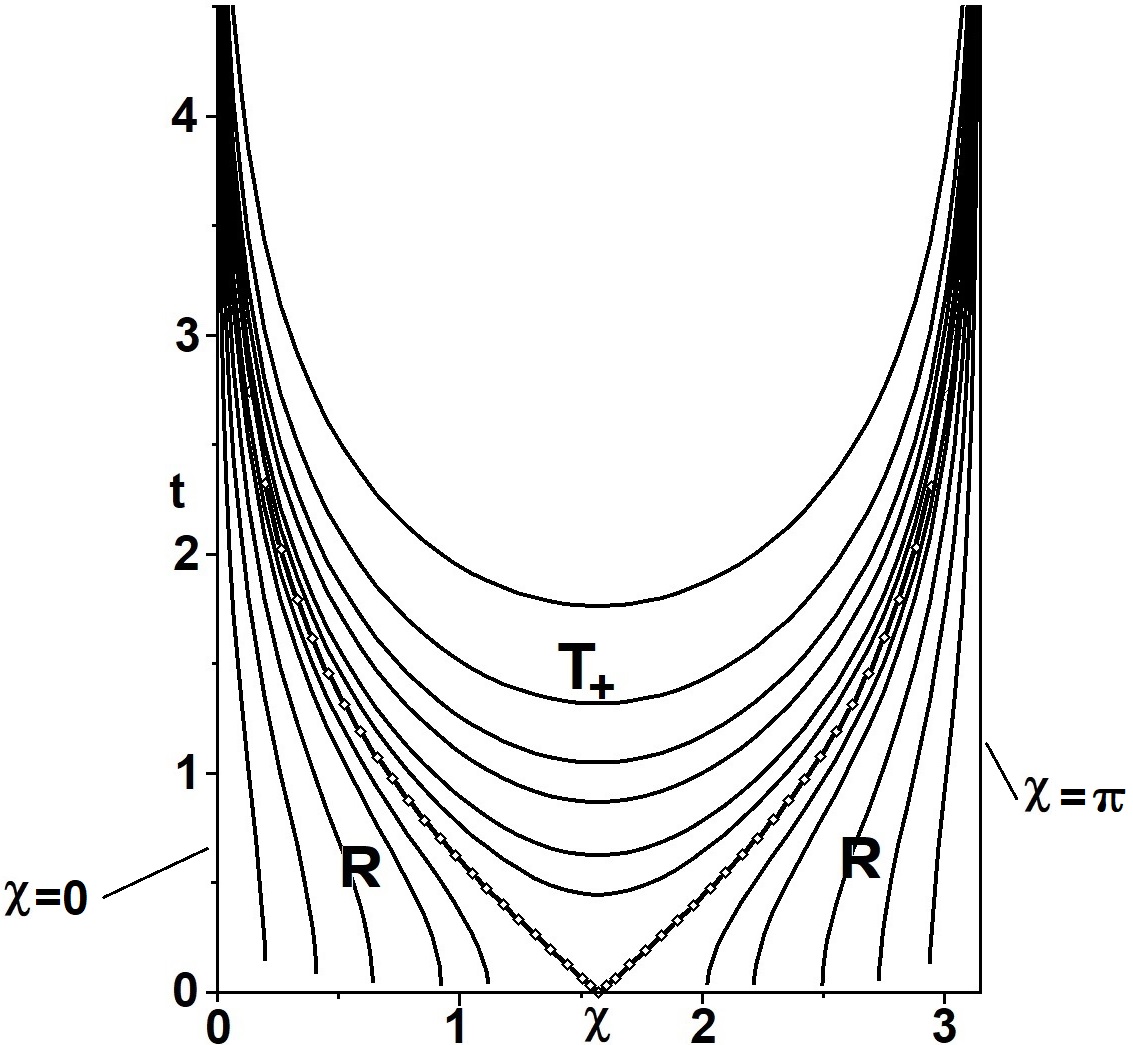} 
      \caption{Reference frame Eq. (\ref{DSframe}) in de-Sitter's universe.
                    Chainlike lines show apparent horizons, solid lines correspond
                    to world lines ($g_{22}=const$). $R$- and $T_+$-regions are shown.}
  \label{Figure_2}
\end{figure}

We use the dimensionless expressions for the length equal to the ratio of the length $\ell$ 
to $a_0$, i.e. $n = \ell/a_0$. In this notation the expression $g_{22} = \text{const}$ can be written as: 
$\cosh^2 t\sin^2 \chi = n^2$ or
\begin{equation}
        t = \arccosh\left(\cfrac{n}{\sin\chi} \right)\,,
        \label{t_func_n}
\end{equation}
and Eq. (\ref{t_coord}) can be rewritten in the form
\begin{equation}
        t = \arccosh\left(\cfrac{1}{\sin\chi} \right)\,.
        \label{t_func_1}
\end{equation}
Figure \ref{Figure_2} shows the graphs of functions (\ref{t_func_n}) and (\ref{t_func_1}) in coordinates $t$,$\chi$. 
Figure \ref{Figure_1}(a) depicts the evolution of an expanding wormhole system, while Fig.~\ref{Figure_2} 
represents a similar evolution of the de~Sitter model. When comparing 
Fig.~\ref{Figure_1}(a) and Fig.~\ref{Figure_2}, it should be remembered that in Fig. \ref{Figure_1}(a), drawn in 
$(u, v)$-coordinates, time $t$ flows along the diagonal from bottom left to top right. In Fig.~\ref{Figure_2}, 
represented in the coordinates $(t, \chi)$, time flows from bottom to top.  For comparison, it is useful to rotate 
Fig.~\ref{Figure_1}a counterclockwise by 45 deg.

      The location of the $R-$ and $T_+$- regions and horizons in both figures are qualitatively similar. However, 
the dynamics of the deformation of the comoving spaces, described by the behavior of the world lines of the reference 
systems, are fundamentally different. In $T_+$-regions, the world lines (they are spacelike) are orthogonal to 
the time line. As $t$ grows, the increasing values of $n$ describe in both figures the expansion along the direction 
orthogonal to the radius. The behavior of world lines in the $R$-areas is different in both figures. In Fig.~\ref{Figure_2}, 
as $t$ increases, they deviate from the center line and from their neighbors. This corresponds to expansion along 
the radius. In Fig.~\ref{Figure_1}(a), the lines deviate toward the center line and their neighbors are located closer 
to the center line. This corresponds to contraction along this direction. 

     Thus, the transverse dimension of the wormhole corridor increases with time, while the length of the corridor 
decreases. This is completely different from the isotropic expansion of the de~Sitter model.

\section{Conclusions}

     As we have shown in this work, the disintegration of a WH leads to an expansion of its transverse dimensions --- 
an increase in the size of the throat (and an increase in volume) --- but does not lead to an approach 
to the de Sitter model. The WH remains a WH, although it increases in size. The size of the throat also increases, which is 
important for an outside observer. Thus, the described process can be considered as a toy model of a possible 
expansion of the WH from quantum dimensions in space-time foam, proposed by Wheeler~\cite{Wheeler_1957}, 
to macroscopic dimensions: a process considered in Refs.~\cite{Thorne_2014, Morris_1988}. See also the note 
in Ref. \cite{Shinkai_2002}.\\

\vspace{1cm}

The work is supported by the Project No. 41-2020 of LPI new scientific
groups and  the Foundation for the Development of Theoretical Physics
and Mathematics ``Basis",  Grant No. 19-1-1-46-1. S.R. thanks R.E.~Beresneva, 
O.N.~Sumenkova and O.A.~Kosareva for the opportunity to work fruitfully 
on this problem. 

\section*{APPENDIX: HOMOGENEOUS APPROXIMATION}  

       \label{Uniform_approx}

A small volume of space on the throat in the first approximation can be considered as a part of a flat 
homogeneous space. In the course 
of expansion, the influence of the scalar field on the expansion dynamics decreases, and we can consider the asymptotic behavior of evolution as we do in an empty of matter model (see Ref. \cite{Zeldovich_1975}). The metric 
is written as
\begin{equation}
     ds^2 = -dt^2 + a^2\,dx_1^2 + b^2\left(dx_2^2 + dx_3^2 \right)\,\,,
\end{equation}
\begin{equation}
     a = a_0 t^{-\frac 13}, \qquad   b = b_0 t^{\frac 23}\,.
\end{equation}
The $x_1$ axis is directed along the radial coordinate, the $x_2$ and $x_3$ axes are pointed in the perpendicular 
direction (a special case of Kasner's solution; see Ref.~\cite{Zeldovich_1975}). The volume $V$ variation law is
\begin{equation}
        V \sim t\,.
\end{equation}

Note that in this approximation, we see the compression in the radial direction and expansion in the directions perpendicular to it.
See also  Ref. \cite{Novikova_2010} for the homogeneous approximation.

\bibliography{wh3}

\end{document}